\documentclass[conference]{IEEEtran}
\IEEEoverridecommandlockouts

\usepackage{cite}
\usepackage{amsmath,amssymb,amsfonts}
\usepackage{algorithm}
\usepackage{algorithmic}
\usepackage{graphicx}
\usepackage{textcomp}
\usepackage{xcolor}
\usepackage{epstopdf}
\def\BibTeX{{\rm B\kern-.05em{\sc i\kern-.025em b}\kern-.08em
    T\kern-.1667em\lower.7ex\hbox{E}\kern-.125emX}}

\begin{document}

\title{Multi-Block UAMP Detection for AFDM under Fractional Delay-Doppler Channel\\

\thanks{This work was supported by the National Natural Science Foundation of
China (No. 62071058). Corresponding Author: Kai Niu.

The authors are with Beijing University of Posts and Telecommunications,
Beijing 100876, China (E-mail: niukai@bupt.edu.cn).}
}

\author{\IEEEauthorblockN{1\textsuperscript{st} Jin Xu}
\IEEEauthorblockA{
\textit{Beijing University of} \\ \textit{Posts and Telecommunications}\\
Beijing, China \\
xujinbupt@bupt.edu.cn}
\and
\IEEEauthorblockN{2\textsuperscript{nd} Zijian Liang}
\IEEEauthorblockA{
\textit{Beijing University of} \\ \textit{Posts and Telecommunications}\\
Beijing, China \\
liang1060279345@bupt.edu.cn}
\and
\IEEEauthorblockN{3\textsuperscript{rd} Kai Niu}
\IEEEauthorblockA{
\textit{Beijing University of} \\ \textit{Posts and Telecommunications}\\
Beijing, China \\
niukai@bupt.edu.cn}
\and
}

\maketitle

\begin{abstract}
Affine Frequency Division Multiplexing (AFDM) is considered as a promising solution for next-generation wireless systems due to its satisfactory performance in high-mobility scenarios. By adjusting AFDM parameters to match the multi-path delay and Doppler shift, AFDM can achieve two-dimensional time-frequency diversity gain. However, under fractional delay-Doppler channels, AFDM encounters energy dispersion in the affine domain, which poses significant challenges for signal detection. This paper first investigates the AFDM system model under fractional delay-Doppler channels. To address the energy dispersion in the affine domain, a unitary transformation based approximate message passing (UAMP) algorithm is proposed. The algorithm performs unitary transformations and message passing in the time domain to avoid the energy dispersion issue. Additionally, we implemented block-wise processing to reduce computational complexity. Finally, the  empirical extrinsic information transfer (E-EXIT) chart is used to evaluate iterative detection performance. Simulation results show that UAMP significantly outperforms GAMP under fractional delay-Doppler conditions. 
\end{abstract}

\begin{IEEEkeywords}
AFDM, fractional Delay-Doppler, MB-UAMP, E-EXIT.
\end{IEEEkeywords}

\section{Introduction}
Reliable communication under high mobility conditions is one of the key challenges for next-generation (6G) wireless networks. This scenario applies to various environments such as high-speed rail transport, aerospace communications, and vehicle-to-vehicle communication. However, in high-mobility environments, multi path delay and Doppler shift lead to linear time varying (LTV) channels with doubly selective fading in both time and frequency. This results in severe inter carrier interference (ICI) for Orthogonal Frequency Division Multiplexing (OFDM)\cite{hadani2017orthogonal}, ultimately degrading the performance of demodulation. Therefore, designing waveforms that are robust to time frequency fading in high-mobility scenarios is crucial.

Orthogonal Time Frequency Space (OTFS)\cite{hadani2017orthogonal} modulation has gained recognition for effectively combating doubly selective fading caused by time-varying channels by operating in the delay-Doppler domain. In recent years, researchers have conducted extensive studies on signal detection\cite{OTFS_AMP,OTFS_dec_2018WCNC,OTFS_dec_2020_TVT}, channel estimation\cite{OTFS_CE_2019,OTFS_CE_2022}, and integrated sensing and communication\cite{OTFS_ISAC_2021,OTFS_ISAC_2022} for OTFS, demonstrating its excellent performance in high-mobility scenarios. However, OTFS still suffers from drawbacks such as high modulation and detection complexity, large pilot overhead, and significant multi-user multiplexing costs\cite{raviteja2019embedded} due to its 2D modulation. 

In response to these issues, a new technique called Affine Frequency Division Multiplexing\cite{bemani2021afdm}, based on affine Fourier transforms, has emerged recently. AFDM adjusts the parameters of the affine transform according to the delay-Doppler characteristics of the LTV channel to achieve full diversity gain. Although its signal structure is significantly simplified compared to OTFS, reliable signal transmission still requires the support of efficient and low-complexity detection algorithms. In \cite{AFDM_LMMSE}, a low-complexity MMSE-based detection algorithm was proposed by utilizing the channel characteristics of AFDM. Due to the sparsity of the equivalent channel matrix in AFDM, message passing algorithms can also be effectively applied to signal detection. The Gaussian Approximate Message Passing (GAMP) algorithm proposed in \cite{AFDM_AMP} achieves good detection performance with low complexity under integer delay-Doppler conditions. However, in fractional delay-Doppler channels, the GAMP algorithm suffers severe performance degradation due to channel correlation. Inspired by unitary transform based solutions for addressing channel correlation in OTFS \cite{OTFS_UAMP,OTFS_block_UAMP}, this paper proposes a multi-block unitary transformation based approximate message passing (MB-UAMP) algorithm. The algorithm first performs multi-block segmentation and unitary transformation on the time domain received signal and channel matrix, followed by approximate message passing.

The main contributions of this paper are as follows:
\begin{itemize}
\item We derive the system transmission model for AFDM under fractional delay-Doppler conditions, demonstrating the phenomenon of energy dispersion in the affine domain channel response matrix caused by fractional delay-Doppler. This further illustrates the sensitivity of the AFDM system to fractional delay-Doppler. 
\item To enhance the robustness of AFDM against fractional delay-Doppler channels, we propose a MB-UAMP algorithm. First, the received signal and the channel matrix in the time domain are subjected to multi-block segmentation and unitary transformation. Specifically, the channel matrix is segmented by rows, and any all-zero columns are removed. On this basis, each block is decomposed using SVD, and the received signal is then processed with a unitary transformation accordingly. Subsequently, after performing time domain message passing for each block, the results are transformed to the affine domain using DAFT/IDAFT to complete signal detection.
\item We perform theoretical analysis and performance estimation of the proposed detector using EXIT charts. Moreover, we propose an E-EXIT chart to provide a more precise analysis of the algorithm.
\end{itemize}

The organization of the following sections is as follows. Section II introduces the complete transmission process of the AFDM system and derives the transmission model under fractional delay-Doppler conditions. Section III provides a comprehensive explanation of the MB-UAMP algorithm and Section IV provides a theoretical analysis of its performance gain using E-EXIT charts. Section V analyzes the simulation results, highlighting the performance advantages of MB-UAMP. Finally, Section VI concludes the paper.

\section{AFDM under Fractional Delay-Doppler}
This section first provides a brief explanation of the AFDM transmission system model, followed by an extension of the model for fractional delay-Doppler channels.

The Discrete Affine Fourier Transform (DAFT) and inverse Discrete Affine Fourier Transform (IDAFT) serve as the mathematical foundation of AFDM modulation. For discrete variables ${\mathbf{s}} = \left( {{s_0}, \cdots ,{s_{N - 1}}} \right)$ and ${\mathbf{S}} = \left( {{S_0}, \cdots ,{S_{N - 1}}} \right)$ it is defined as 
\begin{equation}\label{DAFT_equ}
    {\mathbf{S}} = {\mathbf{As}},{\text{   }}{\mathbf{A}} = {{\mathbf{\Lambda }}_{{c_2}}}{\mathbf{F}}{{\mathbf{\Lambda }}_{{c_1}}}
\end{equation}
\begin{equation}\label{IDAFT_equ}
    {\mathbf{s}} = {{\mathbf{A}}^{ - 1}}{\mathbf{S}} = {{\mathbf{A}}^H}{\mathbf{S}},{\text{   }}{{\mathbf{A}}^H} = {\mathbf{\Lambda }}_{{c_1}}^H{{\mathbf{F}}^H}{\mathbf{\Lambda }}_{{c_2}}^H
\end{equation}
where $\mathbf{F}$ denotes the Discrete Fourier Transform (DFT) matrix. ${{\mathbf{\Lambda }}_{{c}}}$ denotes the affine transform matrix with
\begin{equation}\label{Lambda_equ}
    {{\mathbf{\Lambda }}_c} = diag\left( {{e^{ - j2\pi c{n^2}}},n = 0,1, \cdots ,N - 1} \right).
\end{equation}

Let ${\mathbf{x}} \in {\mathbb{A}^{N \times 1}}$ represent the sequence of QAM modulation symbols in the discrete affine domain. After AFDM modulation, we have
\begin{equation}
    {\mathbf{s}} = {\mathbf{\Lambda }}_{{c_1}}^H{{\mathbf{F}}^H}{\mathbf{\Lambda }}_{{c_2}}^H{\mathbf{x}}
\end{equation}
Assuming a raised cosine roll-off filter is used at the transmitter, the received signal after passing through the fractional delay-Doppler LTV channel can be expressed as
\begin{equation}\label{r=sg+n}
    {r_n} = \sum\limits_{l = 0}^\infty  {{s_{n - l}}{g_n}\left( l \right) + {\omega _n}} 
\end{equation}
where $\omega _n$ denotes the Gaussian noise and 
\begin{equation}
    {g_n}\left( l \right) = \sum\limits_{i = 1}^P {\sum\limits_{l = 0}^{L - 1} {{h_i}{\operatorname{P} _{rc}}\left( {l{T_s} - {\tau _i}} \right){e^{ - j2\pi \left( {{k _i} + {\kappa_i}} \right)}}} } .
\end{equation}
Here, $h_i$, $\tau_i$, $k_i$ and $\kappa_i$ represent the channel response coefficient, delay, normalized integer Doppler shift, and fractional Doppler shift of the $i$-th path, respectively. $P$ denotes the number of multipath components. $P_{rc} \left( \tau\right) $ represents the raised cosine roll-off pulse. $L$ refers to the number of symbols affected by a single symbol due to the delay, which is typically determined by the maximum delay and the tail length of the filter pulse energy, i.e. $L = \left\lceil {{l_{tail}} + {\tau _{\max }}} \right\rceil $.
With proper configurations of $c_1$ and $c_2$, after removing the chirp-periodic prefix, \eqref{r=sg+n} can be rewritten in matrix form as 
\begin{equation}
    {\mathbf{r = }}{{\mathbf{H}}_t}{\mathbf{s + w}}
\end{equation}
where $\mathbf{w}$ denotes the Gaussian noise and the channel matrix ${\mathbf{H}}_t$ is 
\begin{equation}\label{y=Htx}
    {{\mathbf{H}}_t} = \sum\limits_{i = 1}^P {\sum\limits_{l = 0}^{L - 1} {{h_i}{P_{rc}}\left( {l{T_s} - {\tau _i}} \right)} } {{\mathbf{\Delta }}_{\left( {{k _i} + {\kappa_i}} \right)}}{{\mathbf{\Pi }}_l}.
\end{equation}
${{\mathbf{\Pi }}_l}$ and ${{\mathbf{\Delta }}_{\left( {{k _i} + {\kappa_i}} \right)}}$ represent the effects of delay and Doppler shift on the channel, respectively. ${{\mathbf{\Pi }}_l}$ is a matrix formed by circularly shifting the identity matrix to the left by $l$ columns. ${{\mathbf{\Delta }}_{\left( {{k _i} + {\kappa_i}} \right)}}$ is a diagonal matrix with 
$n$-th diagonal element ${e^{ - j2\pi \frac{n}{N}\left( {{k _i} + {\kappa_i}} \right)}}$.

At the receiver, after performing the IDAFT, the received signal $\mathbf{y}$ can be obtained as 
\begin{equation}
    {\mathbf{y}} = \sum\limits_{i = 1}^P {{{\mathbf{H}}_i}{\mathbf{x}} + {\mathbf{\tilde \omega }} }
\end{equation}
with
\begin{equation}\label{Hi}
    {{\mathbf{H}}_i} = {\mathbf{A}}\left( {\sum\limits_{l = 0}^{L - 1} {{h_i}{P_{rc}}\left( {l{T_s} - {\tau _i}} \right)} {{\mathbf{\Delta }}_{\left( {{k_i} + {\kappa_i}} \right)}}{{\mathbf{\Pi }}_l}} \right){{\mathbf{A}}^H}.
\end{equation}
By organizing equations \eqref{DAFT_equ}, \eqref{IDAFT_equ}, \eqref{Lambda_equ} and \eqref{Hi}, the expression for $\mathbf{H}_{i}\left(p,q\right)$ can be obtained as 
\begin{equation}
\begin{gathered}
  {{\mathbf{H}}_i}\left( {p,q} \right) = \sum\limits_{l = 0}^{L - 1} {\frac{{{P_{rc}}\left( {l{T_s} - {\tau _i}} \right)}}{N}} {e^{j\frac{{2\pi }}{N}\left( {N{c_1}{l^2} - ql + N{c_2}\left( {{p^2} - {q^2}} \right)} \right)}} \\ 
   \times \frac{{{e^{ - j2\pi \left( {p - q + {k_i} + {\kappa _i} + 2N{c_1}l} \right)}} - 1}}{{{e^{ - j\frac{{2\pi }}{N}\left( {p - q + {k_i} + {\kappa _i} + 2N{c_1}l} \right)}} - 1}} \\ 
\end{gathered} 
\end{equation}
In the following, we uniformly set the affine parameters $c_1$ to $\frac{{2{k _{\max }} + 1}}{{2N}}$ and $c_2$ to $0$.

Essentially, the effects of fractional delay and Doppler shift in the time domain are energy dispersion and phase rotation. The presence of fractional delay causes the energy of a single path to be dispersed across multiple detectable paths after passing through the filter's impulse response. The larger the fractional delay, the more severe the energy dispersion. For Doppler shift, although it only affects the phase of symbols in the time domain, its impact on the channel in the affine domain after an affine transform exhibits an energy dispersion effect similar to that of fractional delay. Compared to fractional delay, the energy dispersion caused by fractional Doppler is more severe, leading to additional path loss and interference, which hinders the detector's ability to achieve full diversity gain.

\begin{figure}[t]
\centerline{\includegraphics[scale=0.35]{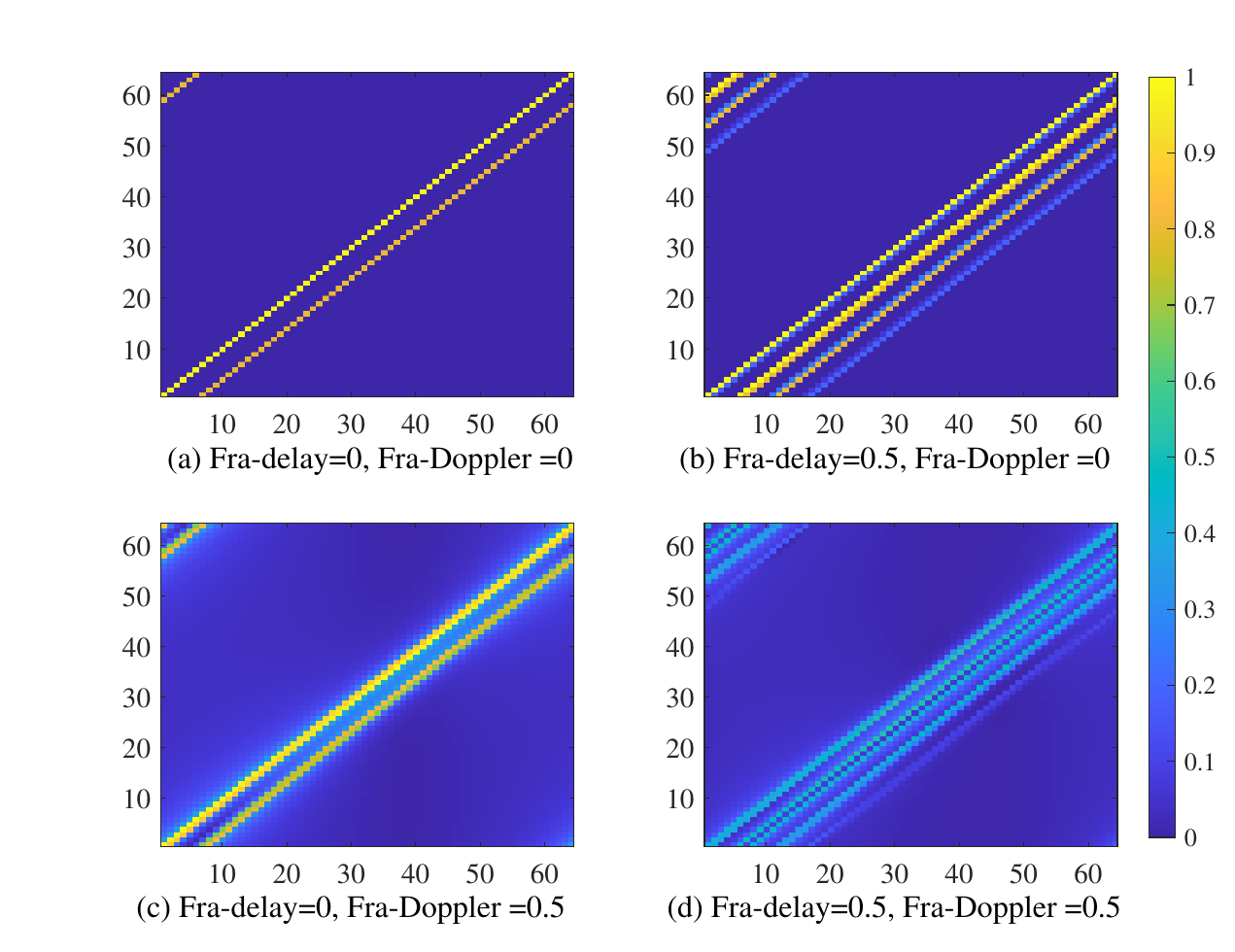}}
\caption{AFDM channel matrix under different delay and Doppler shift.}
\label{AFDM_channel}
\end{figure}

Fig.\ref{AFDM_channel} illustrates the magnitude response of the channel matrix in the affine domain under different fractional delay-Doppler conditions. It can be observed that when the channel experiences fractional delay, the originally two-path channel is dispersed into numerous paths, with the energy of each path being weakened. Conversely, when fractional Doppler is present, the channel energy is dispersed across the entire plane, leading to significant interference between different paths. This implies that it is extremely challenging to reassemble the dispersed energy at the receiver. Therefore, new detection algorithms need to be designed for AFDM under fractional delay-Doppler conditions to combat channel dispersion effects.

\section{Multi-block UAMP Detection}

In this section, we provide a detailed explanation of the multi-block UAMP detection. Fig.3 shows the schematic diagram of the MB-UAMP algorithm. The algorithm is divided into two parts: one part is multi-block unitary transformation processing, and the other part is approximate message passing. We first explained the unitary transformation process with multi-block segmentation, followed by a description of the approximate message passing iterative algorithm based on DAFT/IADFT.

\subsection{Multi-Block Unitary Transformation Processing}

The unitary transformation process has been proven to provide significant gains for message-passing algorithms. However, due to the impact of energy dispersion, the fractional delay-Doppler channel matrix in the affine domain loses its sparsity. Directly applying unitary transformation to it would incur high complexity costs. Considering the sparse banded structure of the time domain matrix, we can first apply multi-block segmentation, followed by unitary transformation, and then perform message passing in the time domain. This approach allows us to leverage the benefits of unitary transformation, without significantly increasing complexity.

\begin{figure}[t]
\centerline{\includegraphics[scale=0.65]{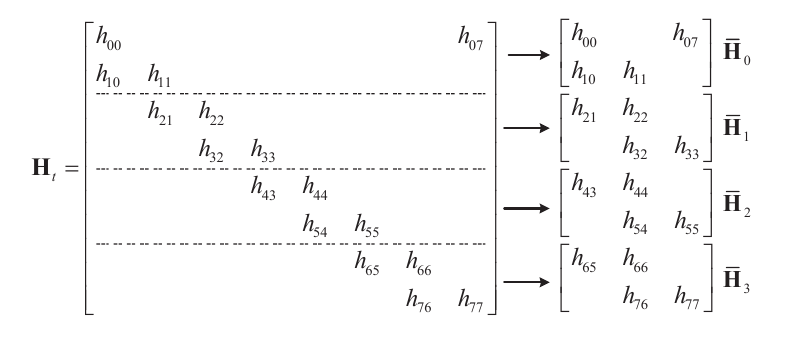}}
\caption{An example of multi-block segmentation of the channel matrix $\mathbf{H}_t$.}
\label{Block_segmentation}
\end{figure}

According to the time domain transmission model (\ref{y=Htx}), we divide the received signal $\mathbf{r}$ and the channel matrix $\mathbf{H}_t$ into $B$ groups, which is called multi-block segmentation. The sequence $\mathbf{r}$ is evenly divided into $B$ blocks, while $\mathbf{H}_t$ is segmented into $B$ blocks according to rows.
\begin{equation}
    {{\mathbf{r}}_b} = {{\mathbf{H}}_b}{\mathbf{s}} + {{\mathbf{w}}_b},{\text{   }}b = 0, \cdots ,B - 1
\end{equation}
where ${{\mathbf{r}}_b} \in {\mathbb{C}^{Q \times 1}}$, ${{\mathbf{H}}_b} \in {\mathbb{C}^{Q \times N}}$, ${{\mathbf{w}}_b} \in {\mathbb{C}^{Q \times 1}}$ and $Q=N/B$. Then, we eliminate the all-zero columns in $\mathbf{H}_b$, resulting in a more concise representation
\begin{equation}
    {{\mathbf{r}}_b} = {{{\mathbf{\bar H}}}_b}{{\mathbf{s}}_b} + {{\mathbf{w}}_b},{\text{   }}b = 0, \cdots ,B - 1
\end{equation}
where ${{\mathbf{s}}_b} \in {\mathbb{C}^{\left(Q+\tau_{max}\right) \times 1}}$, ${{\mathbf{H}}_b} \in {\mathbb{C}^{ Q \times \left(Q+\tau_{max}\right)}}$. Note that ${\mathbf{s}}_b$ is not an equidistant segmentation of $\mathbf{s}$; rather, it is the sequence of transmitted symbols that corresponds to ${\mathbf{r}}_b$ based on 
${\mathbf{H}}_b$. Its index set is defined as $\mathcal{N}_b$. 

Fig.\ref{Block_segmentation} presents an example of multi-block segmentation of a two-path time domain channel matrix, where $N=8$ and $B=4$. One observation is that although the matrix $\mathbf{H}_t$ is very large, only a small portion of it is non-zero. After multi-block segmentation, we can simplify the original matrix into $B$ smaller matrices.

To reduce the correlation within each block, ${{{\mathbf{\bar H}}}_b}$ needs to be decomposed using SVD, i.e., ${{{\mathbf{\bar H}}}_b} = {{\mathbf{U}}_b}{{\mathbf{\Lambda }}_b}{{\mathbf{V}}_b}$. Subsequently, we apply a unitary transformation to the divided received signal, yielding
\begin{equation}
    {{{\mathbf{\tilde r}}}_b} = {\mathbf{U}}_b^H{{\mathbf{r}}_b} = {{\mathbf{\Lambda }}_b}{{\mathbf{V}}_b}{{\mathbf{s}}_b} + {\mathbf{U}}_b^H{{\mathbf{w}}_b} = {{\mathbf{\Phi }}_b}{{\mathbf{s}}_b} + {{{\mathbf{\tilde w}}}_b}.
\end{equation}
At this point, we have completed the multi-block unitary transformation processing.

\subsection{Approximate Message Passing Processing}

\begin{figure*}[t]
\centerline{\includegraphics[scale=0.65]{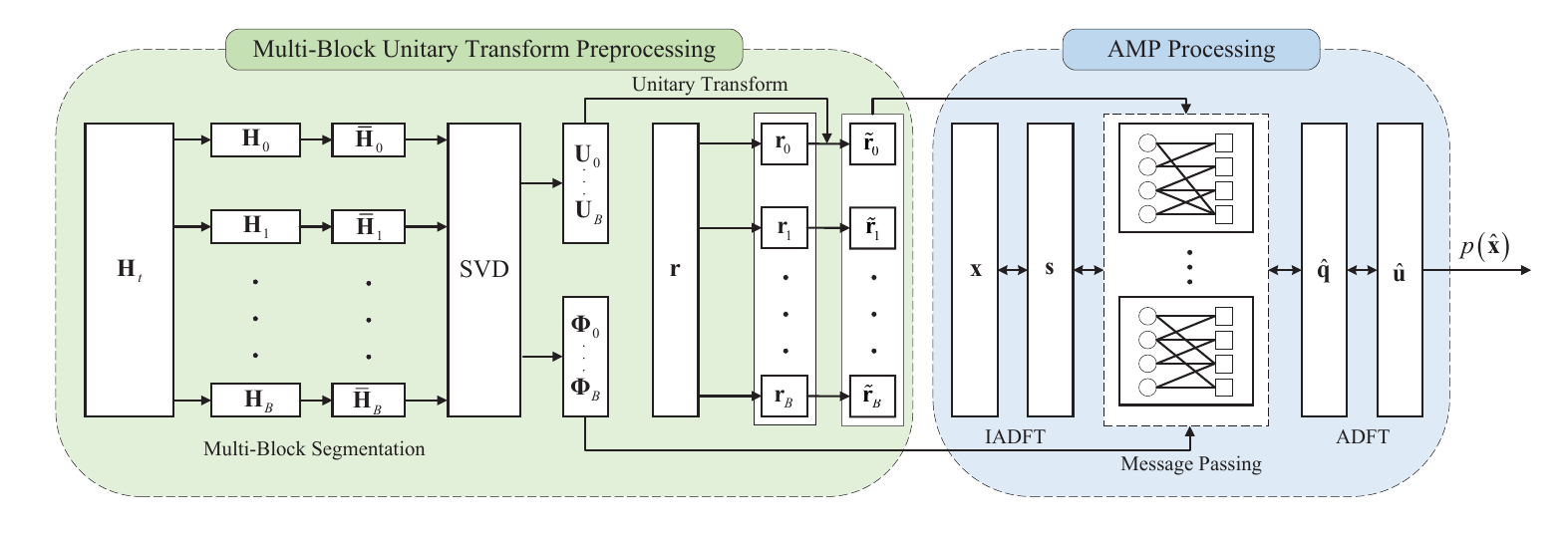}}
\caption{Schematic block diagram of the multi-block UAMP algorithm.}
\label{UAMP_structure}
\end{figure*}

We can map the mean and variance of the affine domain signal $\mathbf{x}$ to the time domain $\mathbf{s}$ through an affine transformation
\begin{equation}\label{AMPbegin}
    {\mathbf{\hat s}} = {{\mathbf{A}}^H}{\mathbf{\hat x}},{\text{   }}{v_s} = {v_x}.
\end{equation}
Based on this, we perform backward message passing processes \cite{OTFS_block_UAMP} as
\begin{equation}
    {{\mathbf{v}}_{{{\mathbf{p}}_b}}} = {\nu _s}{{\mathbf{\lambda }}_b},{\text{ }}{{\mathbf{p}}_b} = {{\mathbf{\Phi }}_b}{{{\mathbf{\hat s}}}_b} - {{\mathbf{v}}_{{{\mathbf{p}}_b}}} \odot {\mathbf{e}}_b^{t - 1}
\end{equation}
\begin{equation}
  \begin{gathered}
  {{\mathbf{v}}_{{{\mathbf{z}}_b}}} = {{\mathbf{v}}_{{{\mathbf{p}}_b}}} \oslash \left( {{\mathbf{1}} + \gamma {{\mathbf{v}}_{{\mathbf{p}_b}}}} \right), \hfill \\
  {{{\mathbf{\hat z}}}_b} = \left( {\gamma {{\mathbf{v}}_{{{\mathbf{p}}_b}}} \odot {{{\mathbf{\tilde r}}}_b} + {{\mathbf{p}}_b}} \right) \oslash \left( {{\mathbf{1}} + \gamma {{\mathbf{v}}_{{{\mathbf{p}}_b}}}} \right) \hfill \\ 
\end{gathered} 
\end{equation}
and forward message passing processes as
\begin{equation}
  \begin{gathered}
  {{\mathbf{v}}_{{{\mathbf{e}}_b}}} = {\mathbf{1}} \oslash \left( {{{\mathbf{v}}_{{{\mathbf{p}}_b}}} + 1/\gamma {\mathbf{1}}} \right), \hfill \\
  {\mathbf{e}}_b^t = {{\mathbf{v}}_{{{\mathbf{e}}_b}}} \odot \left( {{{{\mathbf{\tilde r}}}_b} - {{\mathbf{p}}_b}} \right) \hfill \\ 
\end{gathered} 
\end{equation}
\begin{equation}
  \begin{gathered}
  {{\mathbf{v}}_{{{\mathbf{q}}_b}}} = (Q + {\tau _{\max }}){\mathbf{1}}/\left( {{\mathbf{\lambda }}_b^H{\mathbf{v}}_{{\mathbf{e}}_b}} \right), \hfill \\
  {{{\mathbf{\hat q}}}_b} = {{{\mathbf{\hat s}}}_b} + {{\mathbf{v}}_{{{\mathbf{q}}_b}}} \odot \left( {{\mathbf{\Phi }}_b^H{\mathbf{e}}_b^t} \right) \hfill \\ 
\end{gathered} 
\end{equation}
where $\odot$, $\oslash$ represent element-wise multiplication and division of matrices, respectively. ${{\mathbf{\lambda }}_b} = {{\mathbf{\Lambda }}_b}{\mathbf{\Lambda }}_b^H{\mathbf{1}}$. $\gamma$ is the noise variance. $\mathbf{p}_b$, $\mathbf{z}_b$ and $\mathbf{e}_b$ is the intermediate variables with the variances ${\mathbf{v}}_{\mathbf{p}_b}$, ${\mathbf{v}}_{\mathbf{z}_b}$ and ${\mathbf{v}}_{\mathbf{e}_b}$. ${{\mathbf{\hat q}}}_b$ and  $\mathbf{\hat v}_{\mathbf{\hat q}_b}$ represent the mean and variance of the time domain equivalent detection signal. 
$\mathbf{\hat q}$ can be obtained through the reverse merging operation of the split $\mathbf{s}_b$
\begin{equation}
    {v_{{q_i}}} = {\left( {\sum\limits_{j \in {\mathcal{N}_i}} {v_{{q_{b,j}}}^{ - 1}} } \right)^{ - 1}},{\text{   }}{{\hat q}_i} = {v_{{q_i}}}\sum\limits_{j \in {\mathcal{N}_i}} {{{\hat q}_{b,i}}v_{{q_{b,j}}}^{ - 1}} 
\end{equation}
where $\mathcal{N}_i$ denotes the index set of the $i$-th symbol.
Then, $\mathbf{\hat q}$ is mapped back to the affine domain via ADFT
\begin{equation}
    {\mathbf{\hat u}} = {\mathbf{A\hat q}},{\text{   }}{v_u} = \frac{1}{N}\sum\nolimits_{i = 1}^N {{v_{{q_i}}}} 
\end{equation}
Finally, based on normalization and probability calculations, the mean $\hat{x}$ and variance $v_x$ of the transmitted symbol $x$ can be obtained as
\begin{equation}
    {{\hat x}_i} = \sum\limits_{a = 1}^{\left| \mathbb{A} \right|} {{\alpha _a}{\beta _{i,a}}} ,{\text{   }}{v_{{x_i}}} = \sum\limits_{a = 1}^{\left| \mathbb{A} \right|} {{{\left| {{\alpha _a} - {{\hat x}_i}} \right|}^2}{\beta _{i,a}}} 
\end{equation}
where
\begin{equation}
    \begin{gathered}
  {\xi _{i,a}} = P\left( {{x_i} = {\alpha _a}} \right)\exp \left( { - v_u^{ - 1}{{\left| {{\alpha _a} - {{\hat u}_i}} \right|}^2}} \right), \\ 
  {\beta _{i,a}} = {\xi _{i,a}}/\sum\limits_{a = 1}^{\left| \mathbb{A} \right|} {{\xi _{i,a}}}  \\ 
\end{gathered} 
\end{equation}

At this point, one iteration of message passing is complete. $p\left( \hat x \right)$ can be computed across ${\hat x}$ and $v_x$.

Finally, we analyze the complexity of the multi-block UAMP. It is observed that the complexity of the unitary transform is $\mathcal{O}\left( BQ^3\right)$ and it is performed only once in the algorithm. The primary source of complexity in the iterations comes from matrix multiplication, with each iteration having a complexity of $\mathcal{O}\left( BQ^2\right) +
\mathcal{O}\left( BQ\left|\mathbb{A}\right|\right)$. Thus, the overall complexity of multi-block UAMP is $\mathcal{O}\left( BQ^3\right) + \mathcal{O}\left( n_{ite}BQ^2\right) +
\mathcal{O}\left( n_{ite}BQ\left|\mathbb{A}\right|\right))$ . In contrast, the complexity of the Low-complex MMSE algorithm in \cite{AFDM_LMMSE} is $\mathcal{O}\left( n_{ite} \left(2K^2+1\right)\left(N-M\right) \right)$, where $M =\left( {{l_{\max }} + 1} \right)\left( {2{\alpha _{\max }} + 1} \right)$ and $K$ represents the number of non-zero elements in each column of the channel matrix. The complexity of the GAMP algorithm proposed in \cite{AFDM_AMP} is $\mathcal{O}\left( n_{ite}NK\left|\mathbb{A}\right|\right)$. It should be noted that due to the presence of fractional delay and Doppler, the sparsity of the affine domain channel is significantly reduced, thus amplifying the impact of $L$ on complexity. Additionally, as will be seen in subsequent analyses, MB-UAMP also shows an advantage in terms of iteration count. Finally, because of the block processing, MB-UAMP is more suitable for parallel computation, which facilitates its implementation.

\section{Theoretical Performance Analysis}

In this section, we will conduct a theoretical analysis of the convergence of MB-UAMP and GAMP. For iterative detection algorithms, the EXIT chart is a crucial tool for theoretical analysis. By statistically calculating the mutual information between nodes, the EXIT chart can intuitively illustrate the algorithm's convergence process during iterations. Additionally, by providing the prior probabilities of each node, the theoretical value of the mutual information transmitted during a single iteration can also be computed. For an iterative message passing detector, the mutual information between the probabilities of the detector’s input nodes and the transmitted signal $x$ is called prior information ${I_A} = I\left( {{\mathbf{x}};{{\mathbf{L}}_A}} \right)$, while the mutual information between the output probabilities and $x$ at each iteration is referred to extrinsic mutual information ${I_E} = I\left( {{\mathbf{x}};{{\mathbf{L}}_E}} \right)$. $\mathbf{L}$ denotes the log-likelihood ratio (LLR) sequence of transform bits. When the modulation scheme is determined, the LLR and $p(x)$ can be converted into each other. Therefore, the formula for calculating the mutual information at each iteration is
\begin{equation}\label{IxIu}
I_i^{\left( t \right)} = \sum\limits_{k = 1}^{\left| \mathbb{A} \right|} {\int_{{{\hat x}_i}} {p\left( {{{\hat x}_i}\left| {{x_i} = {\alpha _k}} \right.} \right)} } \log \frac{{p\left( {{{\hat x}_i}\left| {{x_i} = {\alpha _k}} \right.} \right)}}{{p\left( {{{\hat x}_i}} \right)}}d{\hat x_i}, 
\end{equation}
where $p\left( {{{\hat x}_i}\left| {{x_i} = {\alpha _k}} \right.} \right)$ and $p\left( {{{\hat x}_i}} \right) = \sum\limits_{k = 1}^{\left| \mathbb{A} \right|} {p\left( {{{\hat x}_i}\left| {{x_i} = {\alpha _k}} \right.} \right)} $ denotes the likelihood probability distribution and  the posterior distribution, respectively. Based on the above formula, the mutual information of the nodes during each iteration of the algorithm can be calculated using Monte Carlo simulation, allowing for the construction of the EXIT chart to depict the iteration process.

For plotting theoretical values, traditional detectors with independent nodes can model the LLR for each bit using a Gaussian distribution, i.e., $\mathcal{N}\left( {\mu_e ,2\left| {\mu_e} \right| } \right)$. However, due to the energy dispersion effect of the channel matrix under fractional delay-Doppler conditions, the correlation between nodes in the affine domain is greatly enhanced, violating the independence assumption. In most cases, there is no specific relationship between the mean and variance of the LLR. To address this, an empirical EXIT chart was proposed in \cite{SICUAMP}. After statistically obtaining the mean and variance of the LLR during each iteration, Loess fitting \cite{jacoby2000loess} is used to derive the fitting curve. Then, samples from the curve are used to simulate theoretical values. The sampled LLR is input into the detector, and after the message passing, the mutual information can be calculated according to equation (\ref{IxIu}).

\begin{figure}[t]
\centerline{\includegraphics[scale=0.35]{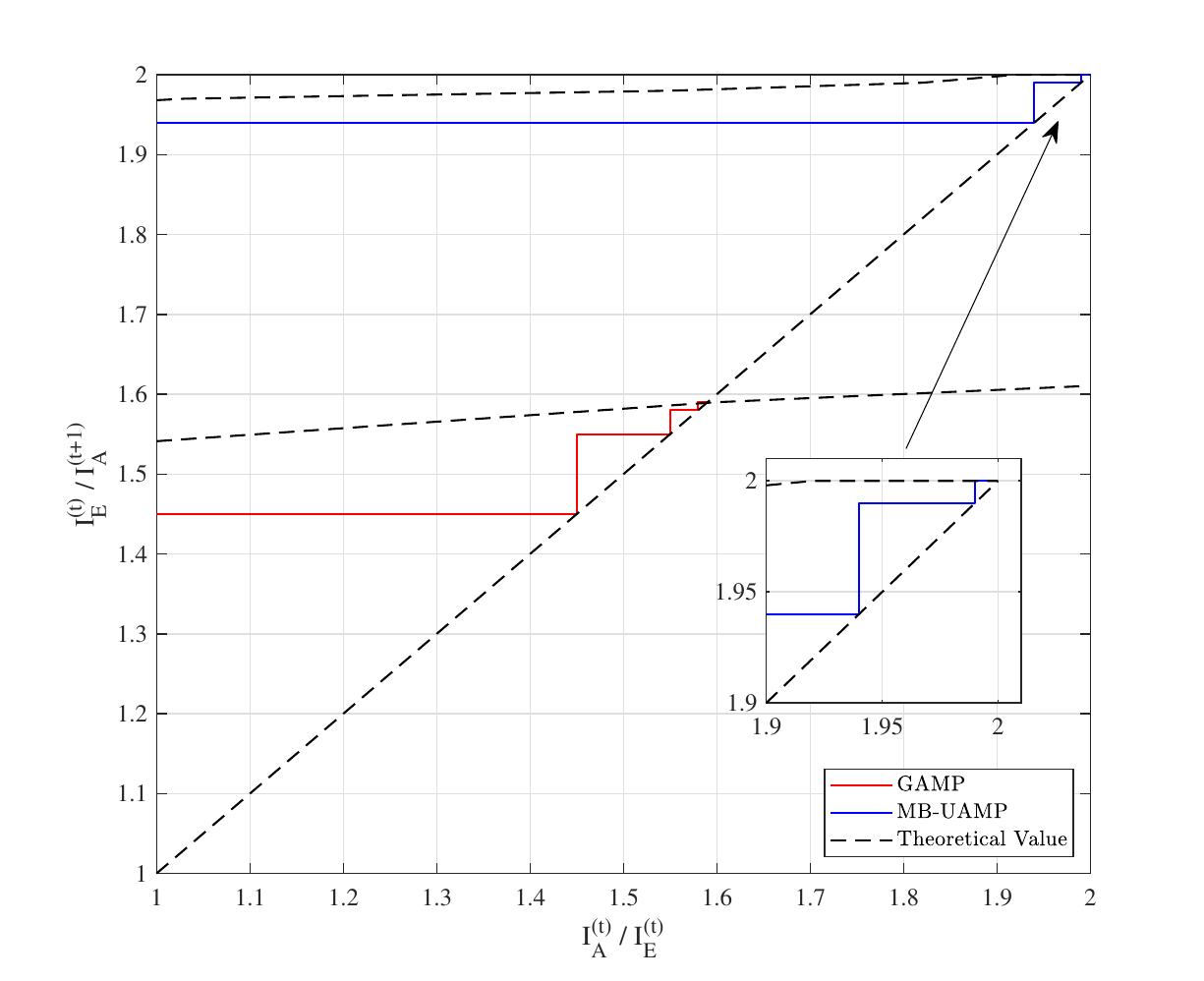}}
\caption{E-EXIT chart for GAMP and MB-UAMP.}
\label{EXIT_chart}
\end{figure}

Fig.\ref{EXIT_chart} presents an example of the E-EXIT chart for MB-UAMP and GAMP. The number of subcarriers is set to $N=128$, with $SNR = 10$ dB. The QPSK modulation is used. To ensure computational stability, a fixed two-path channel is applied in the simulation. The channel's delay, Doppler shift, and channel coefficients are set to $[0.3,1.1]$, $[0.8,1.5]$ and $[1.1+0.6j,0.35+0.21j]$, respectively. The damping factor is set to 0.4.

Due to the absence of an external decoder between iterations, the mutual information output from the previous iteration is identical to the mutual information input for the current iteration, $I_A^{\left( {t + 1} \right)} = I_E^{\left( t \right)}$. As seen in the figure, BM-UAMP's iterative performance significantly surpasses that of GAMP. The theoretical results for GAMP show that as $I_A$ increases, the growth of $I_E$ is very limited. This is primarily due to substantial interference between internal nodes in GAMP, which prevents accurate probability calculations for the nodes. Even when $I_A$ is sufficiently large, the interference still leads to a loss of output mutual information. The intersection of the two theoretical boundaries marks the best possible result for GAMP.

Additionally, the choice of damping factor significantly affects mutual information in GAMP iterations; inappropriate damping can even lead to a decrease in mutual information, indicating poor robustness of GAMP. In contrast, MB-UAMP avoids these issues altogether. Its theoretical bounds suggest it can achieve perfect detection performance due to the sparsity of the cross-domain channel matrix and the independence from the unitary transform. Additionally, practical E-EXIT charts show that MB-UAMP converges faster, typically in 3-4 iterations, while GAMP usually requires 5-6 iterations.

\section{Simulation Results}

In this section, we simulate the performance of the proposed MB-UAMP detector and the GAMP detector \cite{AFDM_AMP} under fractional delay-Doppler channels. Following typical configurations, we conduct tests at a 4 GHz carrier frequency, with the subcarrier spacing to 15 kHz. For one AFDM symbol, we configure 128 subcarriers. The coded modulation scheme adopts QPSK modulation with an LDPC code at a rate of 0.5. The maximum delay and maximum Doppler shift in the digital domain are set to $\tau_{max}=4$ and $(k+\kappa)_{max}=4$. For each path's Doppler shift, we use the Jakes formulation for modeling, i.e. $k_i+\kappa_i=(k+\kappa)_{max}cos\left(\rho_i\right)$. $\rho_i$ is uniformly distributed over $\left[ { - \pi ,\pi } \right]$. The delay index $\tau_i$ is uniformly drawn from $\left({0,\tau_{max}}\right)$ excluding the first path $\tau_0=0$. We focused on testing the system performance under Rayleigh fading channels, where the channel coefficients satisfy 
${h_i} \sim \mathcal{N}\left( {0,{\eta _i}} \right)$ with ${\eta _i} = \exp \left( { - {\tau_i}} \right)/\sum\nolimits_i {\exp \left( { - {\tau_i}} \right)} $.  Additionally, the roll-off factor of
the raised cosine roll-off filter is set to 0.4. Additionally, we simulated the performance of OTFS and OFDM under the same conditions. For OTFS, the number of time slots and subcarriers is set to $M=8$ and $N=16$, respectively, using UAMP detection from \cite{OTFS_block_UAMP}. For OFDM, linear equalization is applied after compensating for the maximum Doppler shift.

\begin{figure}[t]
\centerline{\includegraphics[scale=0.35]{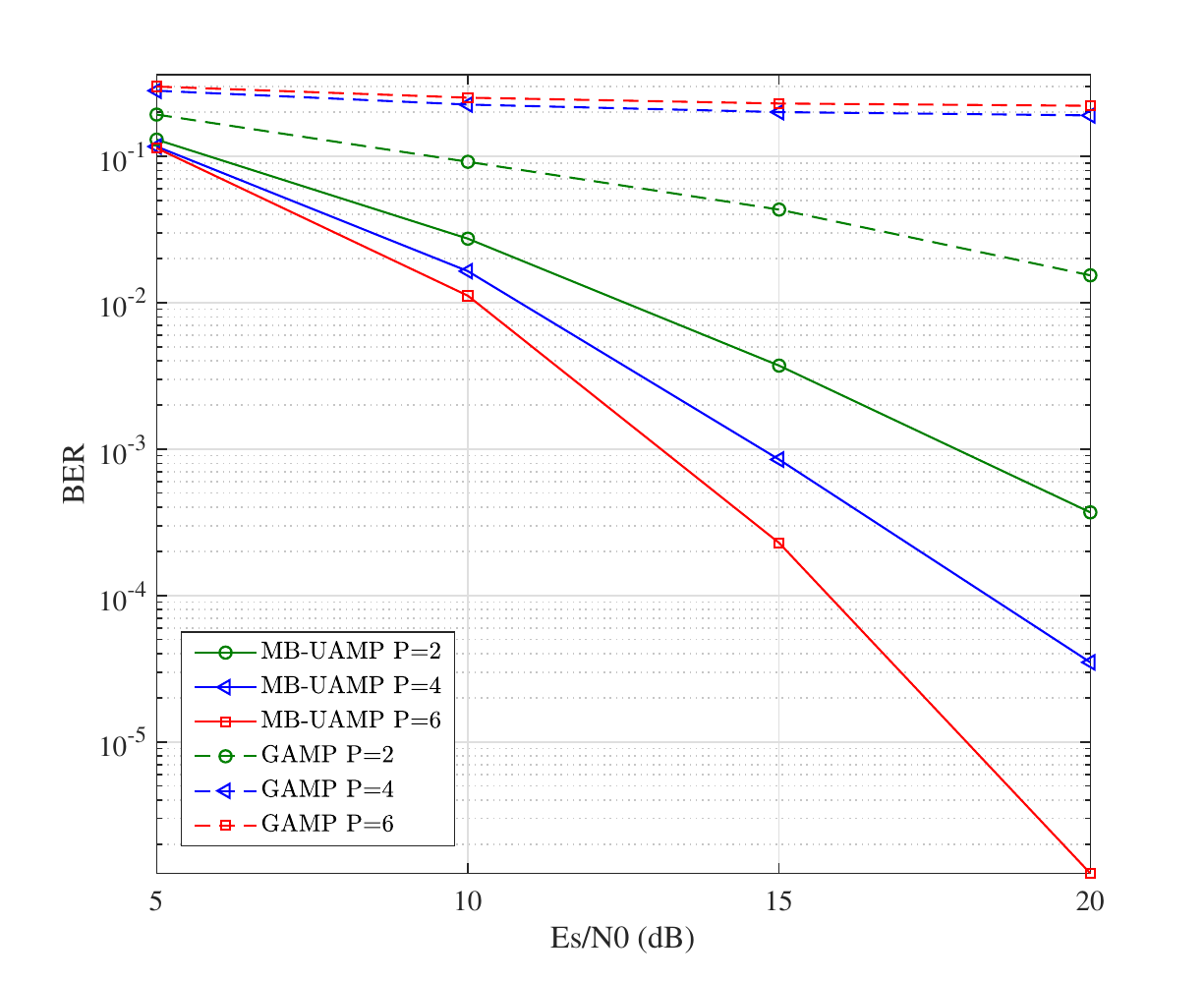}}
\caption{BER performance for multi-block UAMP and GAMP under different number of paths.}
\label{BER}
\end{figure}

Fig. \ref{BER} compares the performance of multi-block UAMP and GAMP algorithms across different path numbers. Under fractional delay-Doppler conditions, MB-UAMP significantly outperforms GAMP, with the advantage growing as the number of paths increases. While interference between paths intensifies, GAMP struggles to manage this interference, leading to a performance decline despite potential diversity gains increasing. In contrast, MB-UAMP effectively mitigates node interference, maximizing the benefits of multipath diversity. Furthermore, GAMP shows a noticeable error floor in high SNR regions, indicating that interference, rather than noise, limits its performance, a problem that MB-UAMP resolves effectively.

\begin{figure}[t]
\centerline{\includegraphics[scale=0.35]{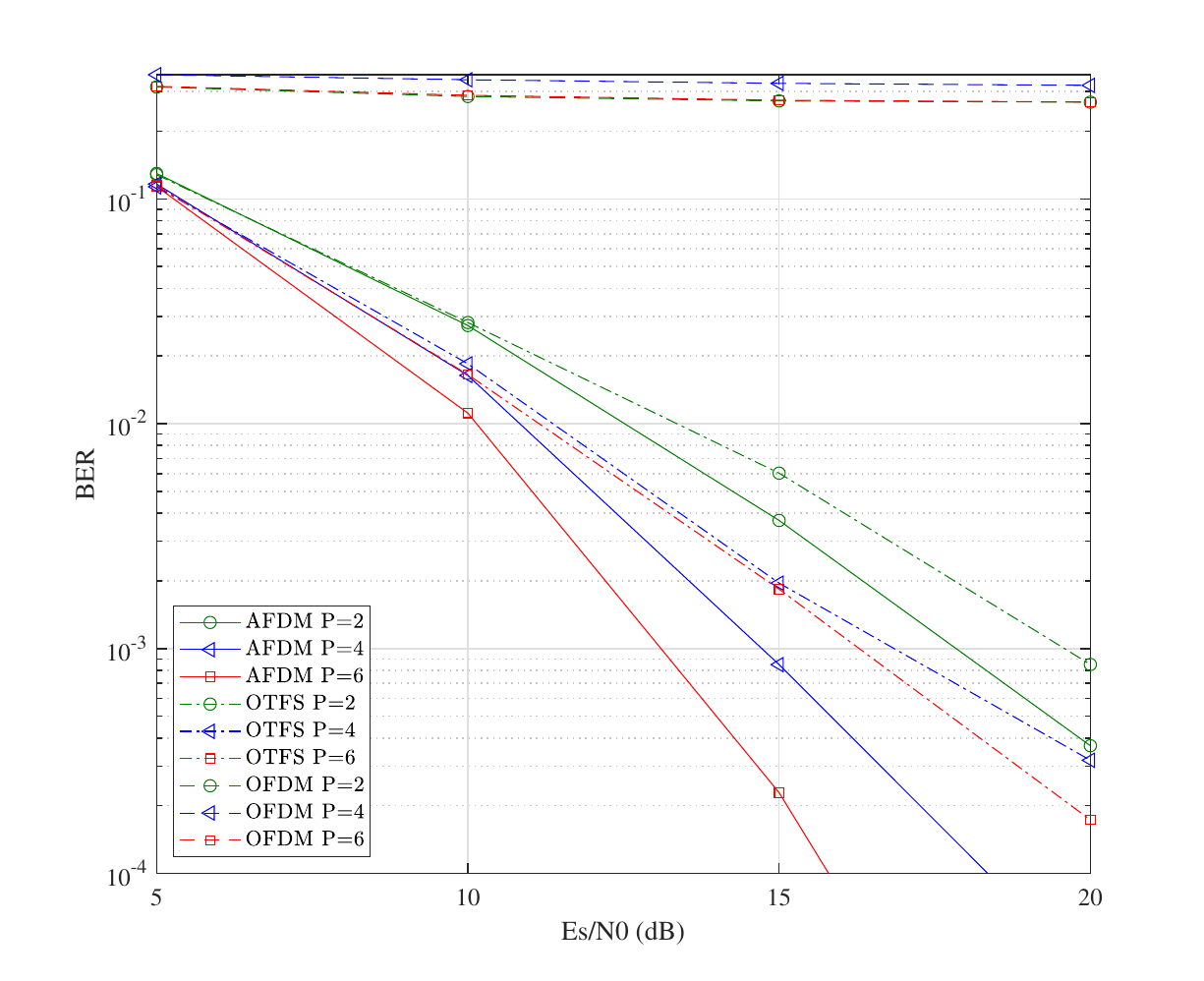}}
\caption{BER performance for AFDM, OTFS and OFDM under different number of paths.}
\label{AFDM_OTFS_OFDM_BER}
\end{figure}

Fig.\ref{AFDM_OTFS_OFDM_BER} shows the simulation results for AFDM, OFDM, and OTFS under different path conditions. OFDM fails completely due to its inability to combat Doppler effects. In contrast, both AFDM and OTFS achieve good detection performance under fractional delay Doppler channels. Notably, AFDM gains 2-3 dB over OTFS at a BER of ${10^{ - 3}}$ using UAMP. This advantage arises from the different symbol carrying methods: AFDM experiences one dimensional interference in the affine domain, while OTFS faces two dimensional interference in the DD domain. Although unitary transformations significantly reduce internal matrix correlation, they cannot fully eliminate it. Thus, AFDM is more robust against interference than OTFS under fractional delay-Doppler.

\section{Conclusion}

This paper analyzes the AFDM system under fractional delay-Doppler channel conditions. A system transmission model for this scenario is presented, followed by identifying the channel energy dispersion issue in the affine domain. To address this, a cross domain MB-UAMP detection algorithm is designed, which mitigates the energy dispersion problem by performing unitary transformation and message passing in the time domain. Subsequently, an empirical EXIT chart is used for theoretical performance analysis of both the MB-UAMP and GAMP algorithms, highlighting the advantages of the MB-UAMP algorithm. Finally, performance simulations of the two algorithms under fractional delay-Doppler conditions are provided, demonstrating the clear performance superiority of the MB-UAMP over the GAMP algorithm.

\bibliographystyle{IEEEtran}
\bibliography{IEEEabrv,reference}

\end{document}